\documentclass[useAMS,usenatbib]{mn2e}
\usepackage{txfonts}
\usepackage{graphicx}

\def\HG{Hi-GAL}
\def\mm{$\mu$m}
\def\30{$\textit{l}=30^{\circ}$}
\def\59{$\textit{l}=59^{\circ}$}
\title[The data reduction pipeline for the Hi-GAL survey]{The data reduction pipeline for the \HG\ survey}

\author[A. Traficante, L. Calzoletti, M. Veneziani et al.]{A. Traficante$^{1,9}$\thanks{E-mail:
alessio.traficante@roma2.infn.it}, L. Calzoletti$^{2,8}$, M. Veneziani$^{3,9}$, B. Ali$^{5}$, G. de Gasperis$^{1}$, A.M. Di Giorgio$^{6}$, 
\newauthor
F. Faustini$^{2}$, D. Ikhenaode$^{4}$, S. Molinari$^{6}$, P. Natoli$^{1,2,7}$, M. Pestalozzi$^{6}$, S. Pezzuto$^{6}$, 
\newauthor
F. Piacentini$^{3}$, L. Piazzo$^{4}$,  G. Polenta$^{2,8}$ and E. Schisano$^{6}$\\
$^{1}$Dipartimento di Fisica, Universit\`a di Roma ``Tor Vergata", Italy\\
$^{2}$ASI Science Data Center, I-00044  Frascati (Rome)\\
$^{3}$Dipartimento di Fisica, Universit\`a di Roma ``La Sapienza", Italy\\
$^{4}$DIET - Dipertimento di Ingegneria dell' Informazione, Elettronica e Telecomunicazioni, Universit\`a di Roma ``La Sapienza", Italy\\
$^{5}$Nasa Herschel Science Center, Caltech, Pasadena, CA\\
$^{6}$INAF-Istituto Fisica Spazio Interplanetario I-00133 Rome\\
$^{7}$INFN Sezione di Tor Vergata, Rome, Italy\\
$^{8}$INAF, Osservatorio Astronomico di Roma, Via Frascati 33, I-00040 Monte Porzio Catone, Italy\\
$^{9}$Spitzer Science Center, Caltech, Pasadena, CA
}

\begin{document}

\date{Submitted}

\pagerange{\pageref{firstpage}--\pageref{lastpage}} \pubyear{2002}

\maketitle

\label{firstpage}

\begin{abstract}
We present the data reduction pipeline for the \HG\ survey. \HG\
is a key project of the Herschel satellite which is mapping the inner
part of the Galactic plane ($|l| \leqslant$ $70^\circ$ and $|b|
\leqslant 1^\circ$), using 2 PACS and 3 SPIRE frequency bands, from
$70$\mm\ to $500$\mm. 
Our pipeline relies only partially on the Herschel Interactive
Standard Environment (HIPE) and features several newly developed
routines to perform data reduction, including
accurate data culling, noise estimation and minimum variance
map-making, the latter performed with the ROMAGAL algorithm, a deep
modification of the ROMA code already tested on cosmological surveys. 
We discuss in depth the properties of the \HG\ Science Demonstration
Phase (SDP) data.
\end{abstract}


%========================================================
\section{Introduction}

The Herschel Space Observatory was launched from Kourou in May 2009 aboard an Ariane 5 rocket. Two of three scientific instruments on the focal plane (PACS and SPIRE) are capable to observe the infrared sky with unprecedented angular
resolution and sensitivity, providing photometric observations in 6
different bands \citep[$70$\mm, $100$\mm, $160$\mm, $250$\mm,
$350$\mm\ and $500$\mm :][and reference therein]{Pilbratt10}.

The PACS photometer is composed of two bolometer arrays: a $64 \times 32$ pixel matrix arranged from 8 monolithic subarrays of $16 \times 16$
pixels each centered on the $70\mu$m and $100\mu$m wavelength (blue and green bands), and a $32 \times 16$ pixel matrix organized in two subarrays for the band centered on $160\mu$m (red band), see \citet{Poglitsch10}.  

SPIRE comprises a three band photometer, operating in spectral bands centered on $250\mu$m, $350\mu$m and $500\mu$m. Every band uses a matrix of germanium bolometers (139, 88 and 43 respectively) coupled to hexagonally packed conical feed horns \citep{Griffin10}.  

In order to handle science data provided by the Herschel instruments, including the data retrieval from the Herschel Science Archive, the data reduction through the standard pipelines and the scientific analysis, an official software environment called HIPE
\citep[Herschel Interactive Processing Environment,][]{HIPE} is available from ESA. 

The raw data provided by the satellite are reduced in HIPE to generate scientific data (so-called Level 2) and
intermediate products of the data reduction process (Level 0 and Level
1 data).

In this paper, we describe the dedicated pipeline created to obtain maps for \HG\ \citep[Herschel Infrared Galactic Plane
Survey,][]{Molinari10_PASP}. Hi-GAL aims to homogeneously cover with observations in 5 contiguous IR bands between 70$\mu$m and 500$\mu$m  a 2 degrees wide stripe of galactic plane between $l=-70^{\circ}$ and $l=70^{\circ}$.

The Galactic plane shows emission varying from point-like sources to large-scale structures and with intensity varying over a wide dynamic range. In this work we show  that  the existing standard reduction strategy (which is based on HIPE version 4.4.0, released in November, 11th 2010) is not optimized to reduce Hi-GAL data and that a dedicated pipeline can enhance the quality of the Level 2 products.

 After Herschel successfully passed the Performance
Verification Phase (PV Phase), two fields of the \HG\ survey were acquired
during the Science Demonstration Phase (SDP): 2x2 square degree areas
of the Galactic plane centered on 30$^{\circ}$ of longitude
(hereafter, \30) and on 59$^{\circ}$ (\59).

We describe the data reduction tools used to obtain high quality maps from SDP data, with the aim to provide a reliable environment for the
Routine Phase (RP) data. The maps provided by our pipeline are successfully used for several works like, e.g. , \citet{Molinari10_special}, \citet{Martin10}, \citet{Peretto10}.

The paper is organized as follows: in Section \ref{sec:higal} we
describe the acquisition strategy for \HG\ data; in Section \ref{sec:preprocessing} we
describe the pre-processing steps of the data reduction pipeline, necessary to prepare data for the map making
and the tools that we have developed to that
purpose. In Section \ref{sec:mapmaking} we describe the ROMAGAL map
making algorithm used in order to estimate the maps. ROMAGAL is used
in place of the MadMap code which is the map making algorithm offered 
in HIPE. The quality of the ROMAGAL maps for both PACS and SPIRE instruments, related to the SDP observations, will be analyzed in Section
\ref{sec:SDP_results}; in Section \ref{sec:Conclusions} we draw our
conclusions. 

%========================================================
\section{The \HG\ acquisition strategy}\label{sec:higal}

\HG\ data are acquired in PACS/SPIRE Parallel mode\footnote{http://herschel.esac.esa.int/Docs/PMODE/html/parallel\_om.html}, in which the same sky
region is observed by moving the satellite at a constant speed of 60$\arcsec$/sec and
acquiring images simultaneously in five photometric bands: $70$\mm\ and $160$\mm\ for PACS and $250$\mm, $350$\mm\ and $500\mu$m for SPIRE.

The whole data acquisition is subdivided in $2^\circ\times2^\circ$ fields of sky
centered on the Galactic plane. Every
\HG\ field is composed of the superposition of two orthogonal AOR (Astronomical Observation Requests).
Each of them is based on a series of consecutive, parallel and partly overlapped scan legs covering $2^\circ\times2^\circ$ square degrees. The scanning strategy adopted for Hi-GAL is fully described in \citep{Molinari10_PASP}. The superposition is performed in order to obtain a suitable data redundancy and for better sampling the instrumental effect like the high-frequency detector response \citep{Molinari10_PASP}.

The acquisition rate for the parallel
mode is 40 Hz for PACS and 10 Hz for SPIRE, although the PACS
data are averaged  on-board for an effective rate of 5\,Hz and 10\,Hz
for the $70$\mm\ and $160$\mm\ array respectively. The implications of the PACS data compression are detailed in Section \ref{sec:signal_striping}. 

\begin{figure}
\centering
\includegraphics[width=4cm]{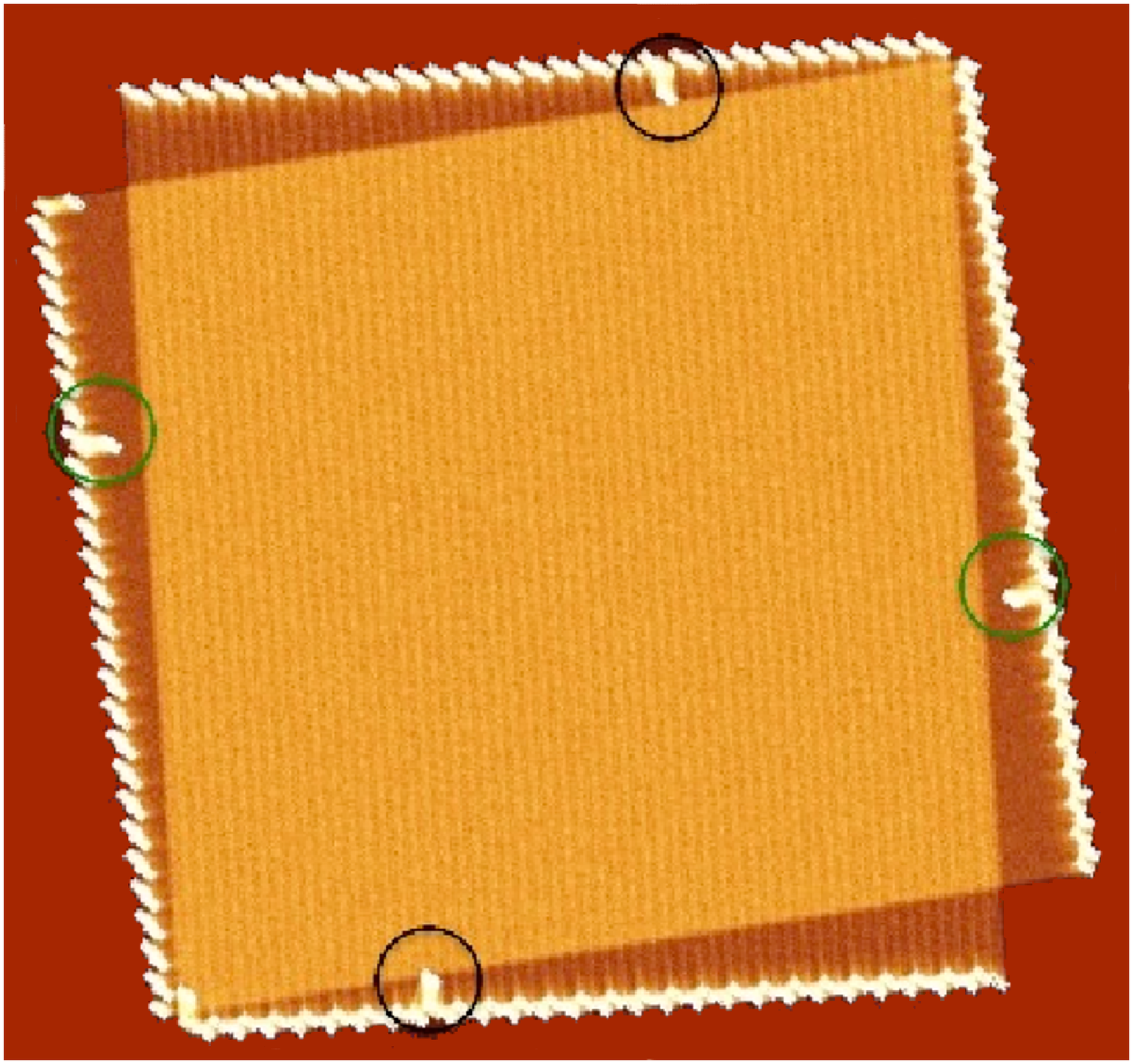}
\includegraphics[width=4cm]{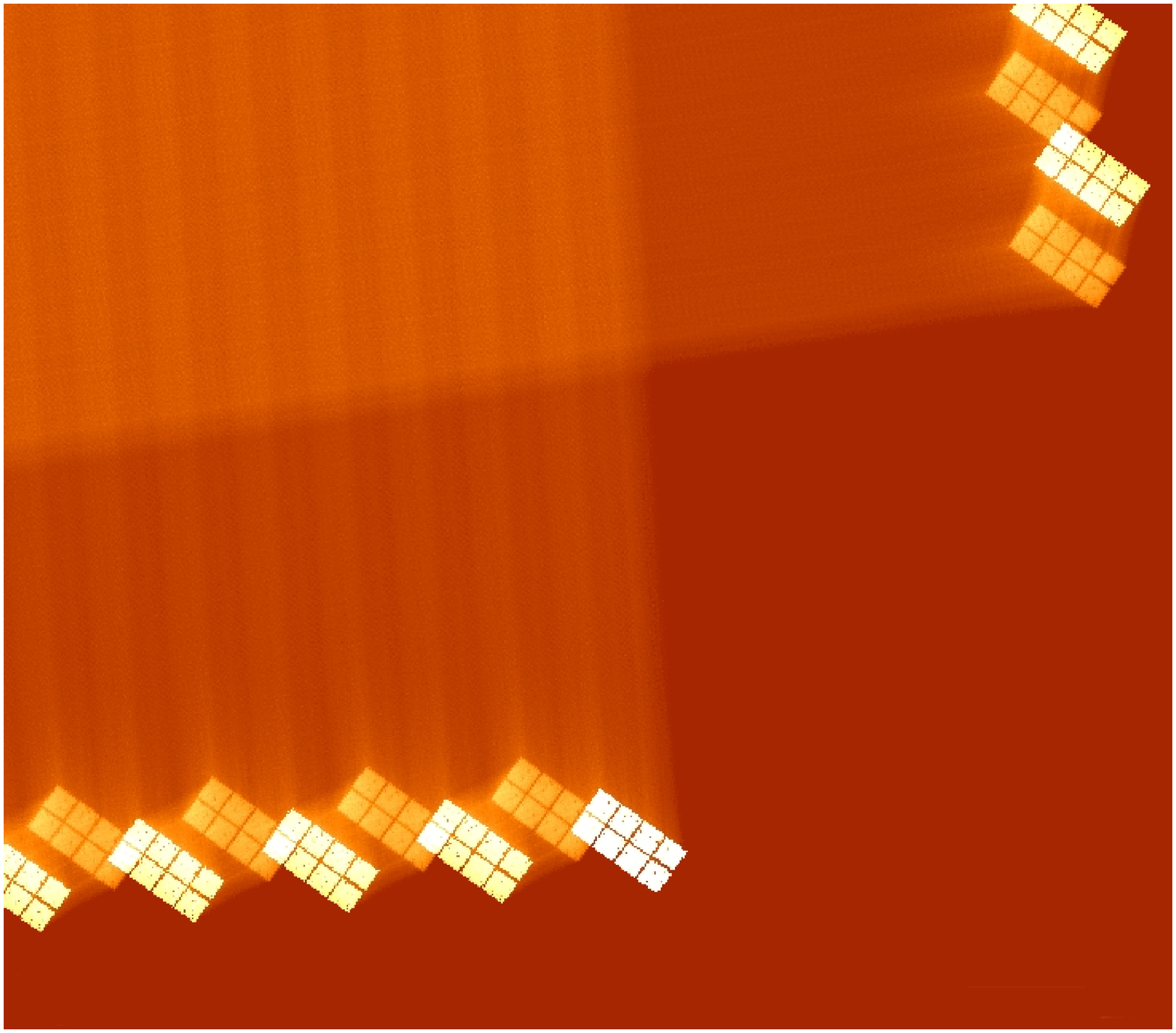}
\caption{Left: the coverage map of PACS blue array. Right: a zoom of the bottom right corner where is clear the
  effect of the superposition from one scan-leg to the next. Black and
  green circles highlight the calibration blocks that, during PV and SDP phase, were observed twice during which the internal calibration sources are observed.} 
\label{fig:coverage}
\end{figure}

An example of the scanning strategy of the \HG\ survey is shown
in Figure \ref{fig:coverage}. The coverage map of the PACS blue array is shown on the left panel and on the right panel we highlight the
superposition of one scan leg to the following ones, by enlarging the
bottom right corner of the left image. Two calibration blocks for each
AOR, during which the instrument observes two internal calibration
sources located on the focal plane, were provided during the SDP
observations. They appear as higher than mean coverage areas and are marked
by black and green circles for the 2 AORs in Figure
\ref{fig:coverage}. Higher coverage zones are also clearly visible in
the slewing region at the end of each scan leg, where the telescope
decelerates and then accelerates before initiating the next scan leg. 

%========================================================
\section{Map making pre-processing}\label{sec:preprocessing}

The aim of the pre-processing is to prepare \HG\ data for
mapmaking. 

While map making is performed
using a Fortran parallel code borrowed from cosmological observations
(see Section \ref{sec:mapmaking}), the preprocessing is done through a series of
IDL and jython tools to be run on the data one after the other. After having
tried to map \HG\ data using the standard tools provided within HIPE,
we decided to develop our own routines that we tailored
specifically for the reduction of data affected by bright and irregular
background, as in the Galactic plane. In fact, high-pass filtering
used in HIPE to cure the long-time drift also removes a large part of
the diffuse Galactic emission. Furthermore, the standard deglitching
embedded in HIPE , the MMT  \citep[Multiresolution Median
  Transform,][]{Starck95}, generates false detections in correspondence
to the very bright sources when we apply this task to the PACS data. On the other hand, the deglitching procedure based on wavelet analysis used by SPIRE does not affect the data, given also the lower spatial resolution compared to the PACS one. We therefore use the HIPE task for SPIRE data only.

Herschel data are stored in subsequent snapshots of the sky acquired
by the entire detector array, called frames. In a first processing
step, the standard modules within HIPE are used to generate Level 1 data for both PACS (except the deglitching task) and SPIRE. Thus, the data are rearranged into one time series per detector pixel, called Time
Ordered Data (TOD), in which the calibrated flux ( Jy\,beam$^{-1}$ for SPIRE and Jy\,sr$^{-1}$ for PACS) and the celestial coordinates are included. 
At the end of this step, TODs are exported outside HIPE in fits format. In the subsequent processing steps, TODs are managed by a series of IDL tools, in order to produce final TODs free
of systematic effects due to electronics and of glitches and corrupted chunks of data due to cosmic
rays. To each TOD a flag file is attached to keep track
of any flagging done during data reduction steps.

\begin{figure}
\centering
\includegraphics[width=4cm]{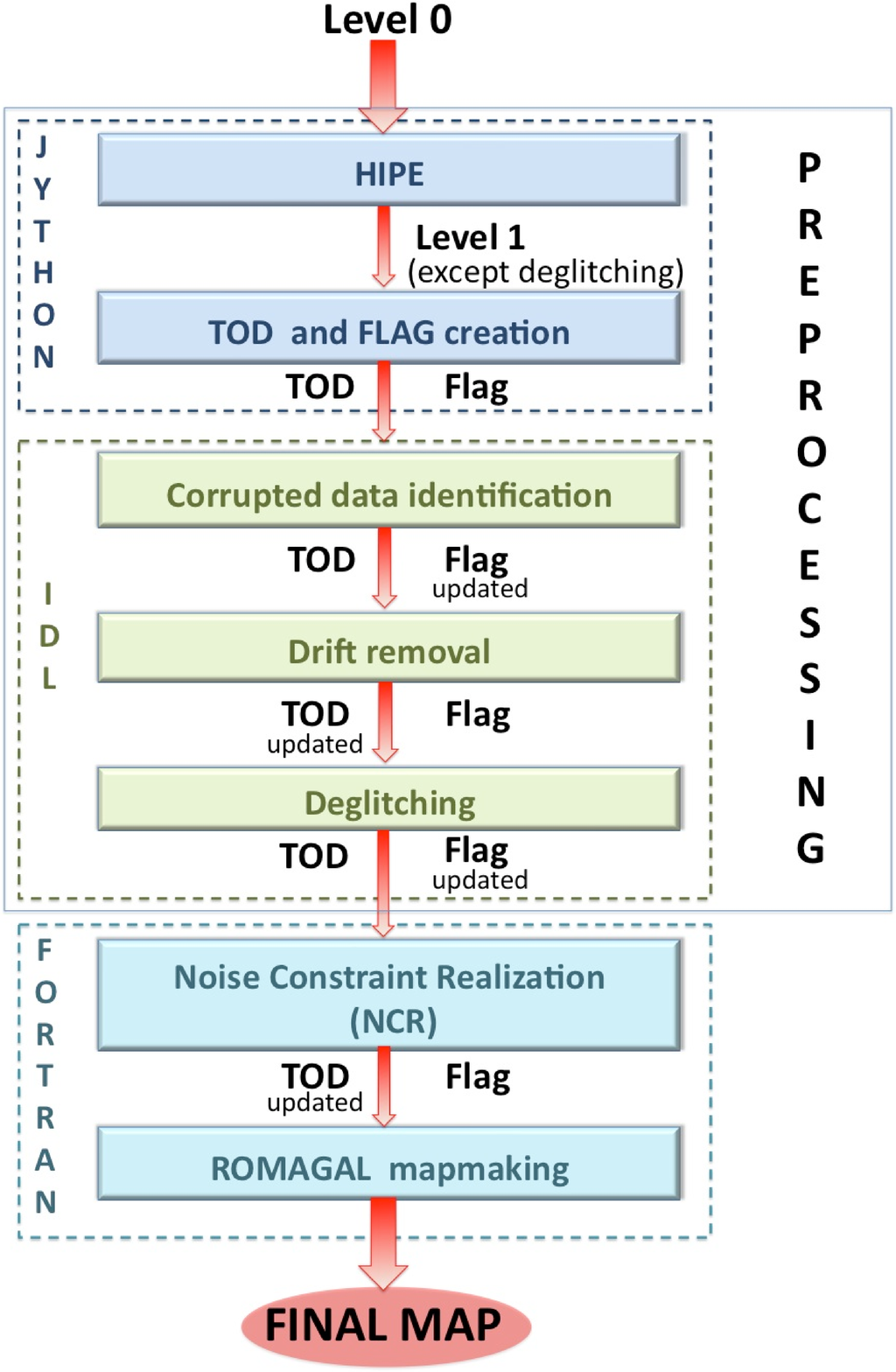}
\caption{Schematic representation of the \HG\ data reduction pipeline} 
\label{fig:higalpipe}
\end{figure}

Preprocessing includes identification of corrupted TODs
(or part of them), drift removal and deglitching.  The following steps will be the Noise Constraint Realization (NCR) and ROMAGAL mapmaking; they will be both described in detail in the next Sections. The summary of the entire pipeline is shown in the diagram in Figure \ref{fig:higalpipe}. 

%========================================================
\subsection{Corrupted TODs}

A TOD can be partially corrupted by the random hiting of
charged particles (cosmic rays) which produce strong and spiky signal
variations called glitches. Two
different effects on the TODs can be identified: the glitch corrupts a single or few consecutive samples, generating
spiky changes along the TOD. This is the most common effect and in
Section \ref{sec:deglitching} we describe how to detect and mask the
data for PACS, as well as to mask any possible residual glitches for SPIRE. Very powerful glitches, on the other hand, %alter the detector
%responsivity, making
make the detector signal unstable for a considerable
interval of time \citep[see, e.g.,][]{Billot10}. This effect depends on the glitch amplitude and on the
bolometers time response. These events affect a much larger part of
the TOD that cannot be used for mapmaking. %The effect of the
%responsivity alteration results in a bias on the bolometer 
%timeline with a low-frequency drift which can involve a considerable
%number of samples, as shown in
%Figure~\ref{fig:bad_bolometer}. 

\begin{figure}
\centering
\includegraphics[width=8cm]{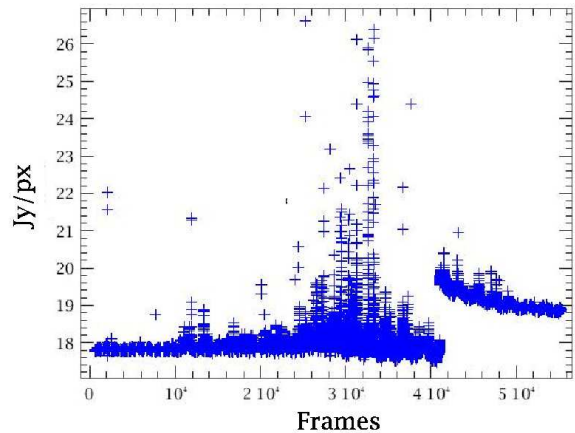}
\caption{Timeline of a PACS blue bolometer. The exponential decay illustrates the change in responsivity after frame 40000 due to the impact of a powerful glitch.}
\label{fig:bad_bolometer}
\end{figure}

Their impact results in a bias on the bolometer 
timeline with a low-frequency drift which can involve a considerable
number of samples, as shown in
Figure~\ref{fig:bad_bolometer}. 

In that Figure, blue crosses represent the
observed timeline of one bolometer of the blue
array. 

Automatic identification of the (partially) corrupted TODs exploits
the first derivative of the TOD to detect extraordinary ``jumps'' in
the signal. In order to determine what portion of the TOD is to be
flagged, the flagging tool exploits the fact that the detector pixels response that have
been hit by a cosmic ray is mostly exponential. Data samples ranging from the
jump to the time sample at which an exponential fit reaches 98\% of
the signal level before the event are identified as bad data and
stored as such in the corresponding flag file. In case the exponential
fit does not reach 98\% of the starting value before the end of the
TOD, then all data starting from the hit will be flagged as is the case
in Figure~\ref{fig:bad_bolometer}. This procedure is applied both in the cases of a changing in responsivity or a detector offset alteration. In the latter, we estimate the fit with an exponent equal to 0. The described procedure was adopted to process both PACS and SPIRE data.

%========================================================
\subsection{Drift removal}

After having identified corrupted data we proceed to the elimination of changes in responsivity over time. The procedures are in principle
identical for PACS and SPIRE data, the only differences account for
different composition of the detector arrays and the data acquisition
of the two instruments. 

The signal in PACS TODs  that exit HIPE does not represent the true sky
but is dominated by the telescope background and the (unknown) zero level of the electronics. The
electronics further introduce significant pixel-to-pixel
offsets. For each TOD, we mitigate the effect of pixel-to-pixel offset by calculating and then subtracting the median level for each pixel from each readout.  This ensures that all pixels have median level equal to 0.  The median value is preferred over mean because the median is a much better representation of the sky+telescope background flux, and is much less sensitive to signal from astrophysical sources.

\begin{figure}
\centering
\includegraphics[width=8cm]{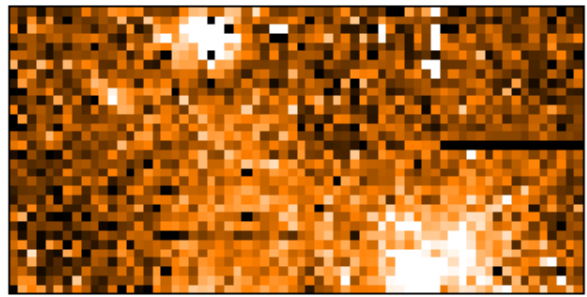}
\caption{\textbf{Blue PACS frame after the median subtraction on each pixel. Diffuse emission and compact source are visible in the frame.}}
\label{fig_drift_2}
\end{figure}

Also, the subtraction of this offset from each TOD does not
alter the signal in the final map, but it
introduces only a global offset constant over the entire area covered
in the observation. However it should be kept in mind that bolometers
are inherently differential detectors which bear no knowledge of an
absolute signal value; besides, any optimized map making methods like
the one we employ (see Section \ref{sec:mapmaking}) produce maps with
an unknown offset value which needs to be independently calibrated. So
it is important to reduce all the bolometers to the same median value,
regardless of its amount. The pixel-to-pixel median subtraction has the effect
seen in Figure~\ref{fig_drift_2}. Diffuse emission and compact sources
are clearly visible in the frame. 

Still, when plotting a detector pixel timeline we see that the signal decreases
systematically from the start to the end of the observation.  This trend is likely due to a combination of small changes in the thermal bath of the array and to small drifts in the electronics. The former affects the entire array, while the latter affects subunits of the detector array (in fact, PACS blue is divided into 8 units, PACS red into 2, electronically independent units \citep{Poglitsch08}). The drift is then a combination of these two effects: drifts of the entire array and drift of a single subunit. These effects are dominant with respect to the $1/f$ noise pattern, which will be described in the next Sections. 

\begin{figure}
\centering
\includegraphics[width=8cm]{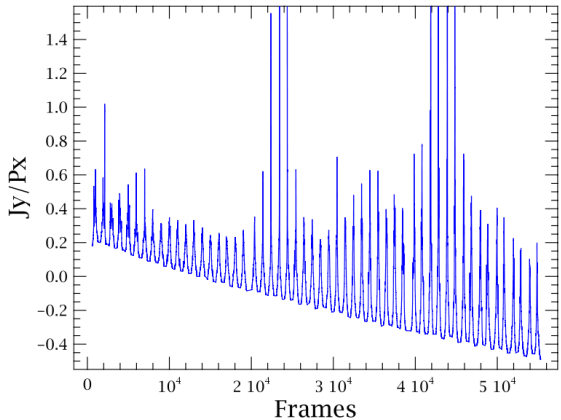}
\caption{Median behavior computed on the whole array for each frame. The
  slow-drift behavior is due to the electronics and the thermal bath.} 
\label{fig_drift_4}
\end{figure}

It is in principle not a trivial task to decide which drift has to be
subtracted first: the drift from the thermal bath (affecting the entire detector
array) or the drift from the readout electronics (affecting sub-arrays
differently)? Ideally
both should be subtracted, if only it were possible to separate each
component accurately, as the net effect  in the data is the sum
of both.

Our methodology for removing the correlated signal drifts (on both the bolometer module/unit level and the array level) is based on tracing the low signal envelope of the unit or array median levels.  In Figure \ref{fig_drift_4}, this envelope is the curve defined by the lowest signal values. and estimated as follows: 

\begin{itemize}
\item[i] We compute the median value of the entire bolometer array/unit for each bolometer readout.  Figure \ref{fig_drift_4} shows one example for the entire array.  
\item[ii] The median values thus obtained are segmented and grouped by scan legs. Each scan leg is composed of $\sim1000$ frames and we observed 54 scan leg for each $2^\circ\times2^\circ$ Hi-GAL field.
\item[iii] For each scan leg we compute the minimum value of the array/unit medians.
\item[iv] The resulting set of minimum median values for all scan legs are fit with a polynomial.
\end{itemize}

The median value for each array/unit readout is chosen because it is closest to the actual sky+telescope background.  However, as clearly seen in Figure \ref{fig_drift_4}, in the presence of strong astrophysical sources the median value is incorrect for our purposes.  The strong sources appear as signal spikes in Figure \ref{fig_drift_4}.  Hence, we take the next step of finding the minimum value from the set of medians belonging to a single scan leg.  The idea is that at some point during the scan leg the median was indeed a true representation of the local sky+telescope and relatively free of source emission.  This step allows us to reject the sources at the expense of degrading our time-resolution to scan-leg duration ($\sim240$ sec).  The polynomial fit allows us to estimate the drift behavior at the same time resolution as the PACS signal (5Hz and 10Hz for 70$\mu$m and160$\mu$m band respectively).  We further note that the correlated signal drift is relatively flat over a single scan leg; hence, the minimum value is not significantly affected by the presence of the monotonic signal drift itself in the first place.

The minimum median method discussed above removes background levels from spatial emission structures that are of the order or larger than the scan legs yet preserves the spatial structures smaller than scan legs.  In essence, information about the absolute calibration zero-point \citep{Miville05} is lost but all spatial structures within our map boundaries are preserved.

\begin{figure}
\centering
\includegraphics[width=8cm]{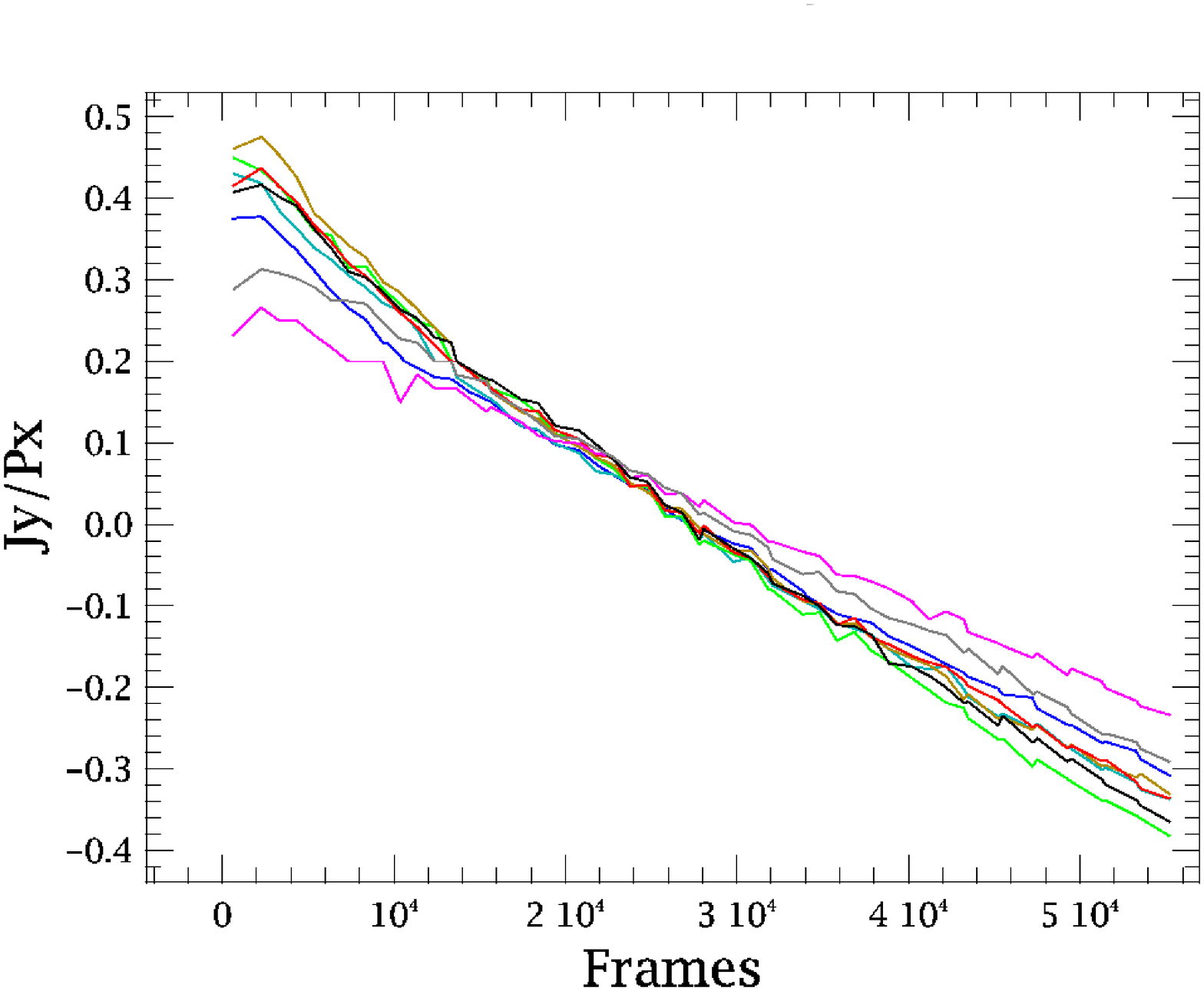}
\caption{Interpolation of the minima of the
  median evaluated on every scan leg. Each curve refers to a PACS subarray. The curves mimic the same behavior with slopes due to the different subarray electronics.} 
\label{fig_drift_6}
\end{figure}

In Figure \ref{fig_drift_6} we reported the minimum median values of each subarray. The common downward trend is due to the (common) thermal bath drift, while the different readout electronics are responsible for the differences in the subarray slopes.

%polynomial fit we have done for interpolating the drift behavior of each subarray. The common downward trend is due to the (common) thermal bath drift, while the different readout electronics are responsible for the differences in the subarray slopes.

We therefore decide to subtract the subarray drifts in order to consider both the thermal bath and the readout electronics behaviors, but separately for each subunit.

\begin{figure}
\centering
\includegraphics[width=8cm]{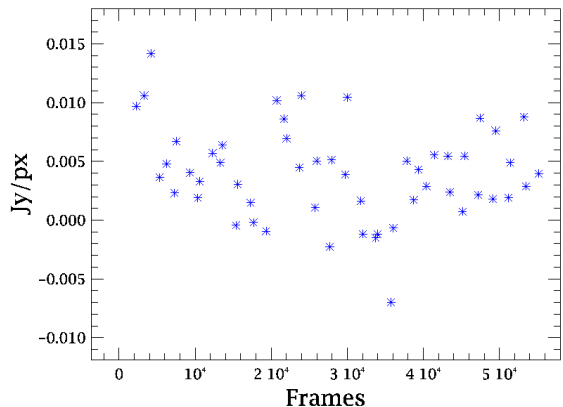}
\caption{Minima of the median evaluated on the whole PACS blue array after the subarray drift
  subtraction. The dispersion is due to the
  intrinsic efficiency of the drifts subtraction tool. There is no
  residual behavior and the scatter is one order of magnitude under
  the value of the initial drift.} 
\label{fig_drift_7}
\end{figure}

Once the individual subarray drift is removed, the remaining dispersion on the whole array is only a residual scatter due to the intrinsic efficiency of the removal tool, as shown in Figure \ref{fig_drift_7}. 

SPIRE array detectors are not divided into subarrays, so every
procedure that has to be run 8 or 2 times for PACS data is only
performed once per SPIRE band. 
SPIRE uses blind bolometers (5 in total) as thermistors to
evaluate the most relevant correlated noise component: the bath
temperature fluctuations. A standard pipeline module uses this
information to perform an effective removal of the common drift
present along the scan observation. HIPE also corrects for the delay between
the detector data and the telescope position along the scan, using an
electrical low-pass filter response correction. But despite these
initial (and very effective) corrections, we apply the drift removal
tool to SPIRE data in the same way as for PACS
data: we fit a polynomial to the minimum of the median of each scan
leg (calculated over the entire detector array), that we then subtract
to all TODs. Experience on the data shows that a residual long time
drift is often present in SPIRE data.  

Finally, when removing drifts it is important to know how the
observational scans are oriented. In fact, as the Galactic plane is
very bright, scans across the plane will give rise to an increase of
signal, on top of the general drift. On the other hand, when the scan
is almost parallel to the plane of the Galaxy, the signal can
dominated by its bright emission, also on the evaluation of the minima of the median. 

\textit{In this case, the curve to fit is estimated only in the scan legs where the median is not affected by the signal.}

Since the procedure is not automatic, care has to used when choosing what
polynomial to fit and subtract from the data, in order not to
eliminate genuine signal from the sky. From our experience the best choice is a first or a second degree polynomial, depending on the signal behavior observed. 

Higher polynomial degree can be necessary when part of the drift has to be extrapolated in order to avoid signal contamination.

\begin{figure}
\centering
\includegraphics[width=8cm]{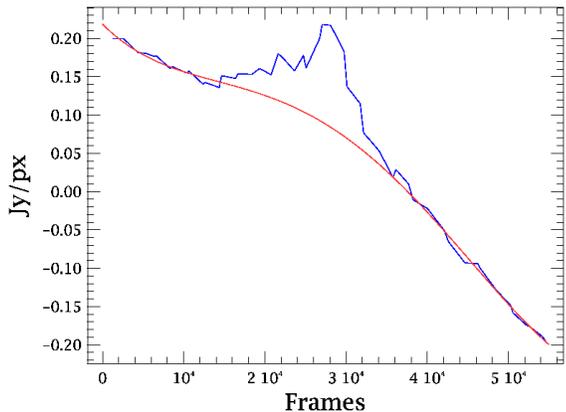}
\caption{Blue curves: interpolation of the minima of the median for one subarray
  of PACS blue channel when the scan direction is almost parallel to the Galactic plane. Red line: the fit that we
  choose to evaluate the drift without considering the minima affected
  by the Galactic emission (central bump).} 
\label{fig:drift_galassia}
\end{figure}

An example for one subarray is shown in Figure~\ref{fig:drift_galassia}. 

Each scan leg is bigger then the overlapping region of the two scan directions for the scanning strategy adopted by Hi-GAL (see Figure \ref{fig:coverage}). Since the Hi-GAL fields are squared regions, the slowly-traversed direction of the AOR within the overlapping region have a length comparable with the scan leg. Thus, we assume that even if there is a signal gradient along the slowly-traversed direction of the AOR, it is not filtered out by the array medians subtraction.

%========================================================
\subsection{Deglitching}\label{sec:deglitching}

To remove outliers in the time series of the bolometers we exploit the spatial redundancy provided by the telescope movement which ensures that each sky pixel of the final map is observed with different bolometers. Outliers detection is done with the standard sigma-clipping algorithm: given a sample of $N$ values, first estimates for the mean and for the standard deviations are derived; then all the values that differ from the mean by more than $n$ standard deviations are considered outliers and removed. 

For this algorithm the choice of $n$, the parameter that defines the threshold above which a value is considered an outlier, is usually arbitrary: a certain $n$ is chosen, very often equal to 3, without any statistical justification. Recently \citet{Pezzuto11} has derived a formula that starting from the properties of the error function for a Gaussian distribution and exploiting the discreteness of the number of available measures, relates $n$ to the size of the sample. The formula is

\begin{equation}
n=-0.569+\sqrt{-0.072+4.99\log(N)}\label{nfromN}
\end{equation}

As a consequence, in the central region of the map, where the coverage (and so $n$) is high, the number of standard deviation is larger than in the outskirt of the map where the coverage is low. For instance, if a sky pixel has been observed with 40 bolometers, the above formula gives $n=2.25$; so, once we have estimated the mean $m$ and the standard deviations $\sigma$, all the values $x_i$ such that ABS$(x_i-m)>2.25\sigma$ are flagged as outliers. If a pixel has been observed with 20 bolometers the threshold lowers to 1.96$\sigma$.

This procedure is automatically iterated until outliers are no longer found. However, the procedure converges within 1 iteration in $\sim$98\% of the cases in which we have applied the analysis.

The outliers detection is done in this way for both instruments, however for SPIRE, as explained before, we also make use of the standard deglitching algorithm (wavelet based) implemented in the official pipeline. But we found some weak glitches left in the SPIRE TODs so that we decided to run our deglitching algorithm also on SPIRE data.

The number of glitches found is on the average about 15\%, a value which is likely larger than the real percentage. For PACS we are now working on a different way to associate each bolometer to the sky pixels, taking into account the finite size of the sky pixels. For the first test cases we run, the percentage of detected glitches is now around 5-6\%.

%======================================================================
\section{The ROMAGAL Map Making algorithm}\label{sec:mapmaking}

The ROMAGAL algorithm is based on a Generalized Least Square (GLS)
approach \citep{Lupton93}. Since the TOD is a linear combination of
signal and noise, we can model our dataset $\textbf{d}_{k}$ for each
detector $k$ as \citep{Wright96}: 

\begin{equation}\label{eq:tod}
\textbf{d}_{k}=P\textbf{m}+\textbf{n}_{k}
\end{equation}

\noindent where $P$ is the pointing matrix, which associates to every
sample of the timeline a direction in the sky, $\textbf{m}$ is our map estimator of the
``true" sky and $\textbf{n}_{k}$ is the noise
vector.  

The observed sky, $P\textbf{m}$, is the map estimator of the ``true sky" convolved with the instrumental
transfer function and the optical beam. However, in case of circularly
symmetric beam profile, $\mathbf{m}$ is a beam smeared, pixelised
image of the sky.

In this case the pointing matrix has only a non-zero
entry per row corresponding to the sky pixel observed at a given
time. Since the beam profiles for PACS  \citep{Poglitsch08} and SPIRE
\citep{Griffin09} are only weakly asymmetric we can put ourselves in
this simple case. Note that the transpose of the $P$ operator performs
a data binning (without averaging) into the sky pixels. 

Equation~\ref{eq:tod} holds only if the noise vector of each detector $\textbf{n}_{k}$ is
composed of statistical random noise, with Gaussian distribution and
null average. All the relevant systematic effects (offset, glitches) have then to be removed with an accurate data preprocessing
before map production, as explained in Section
\ref{sec:preprocessing}.

The formalism can be easily extended in the case of multidetector analysis. In this case the vector $\textbf{d}$ contains the data relative to each detector. Rather, one has to take care to upgrade also the noise vector $\textbf{n}$, accordingly with the correct noise value for each detector.

The GLS algorithm produces minimum noise variance sky maps. Noise
properties for each detector  have to be previously estimated and provided in input to
the algorithm as described in Section \ref{sec:noise_filters}.  

The GLS estimate for the sky, $\tilde\textbf{m}$, is \citep{Natoli01}
\begin{equation}\label{eq:GLS}
\tilde\textbf{m}=(P^{T}\textbf{N}^{-1}P)^{-1}P^{T}\textbf{N}^{-1}\textbf{d}
\end{equation}
\noindent where $\textbf{N}=\langle \textbf{nn}^{T}\rangle$ is the
noise covariance matrix, which takes into account noise time
correlation between different samples. Such correlation is
particularly high at low frequencies because of the $1/f$ (or long
memory) noise. In case of uncorrelated noise (or white noise) the $\textbf{N}$
matrix becomes diagonal and the problem is greatly simplified.  If we
further assume stationary uncorrelated noise, Equation \ref{eq:GLS} reduces
to: 
\begin{equation}\label{eq:naive}
\tilde\textbf{m}=(P^{T}P)^{-1}P^{T}\textbf{d} .
\end{equation}
$P^{T}P$ is the number of observations of a pixel of the map, so we
are averaging the different TOD values into that pixels assigning the
same weight to each sample. We will refer to this map estimate as ``naive" or ``binned" in the following.  

When non negligible noise correlation is present, as in the case of
PACS \citep{Poglitsch08} and SPIRE \citep{Schulz08},  Equation~\ref{eq:GLS} must be solved. This is a challenging
computational task since it requires, in principle, the inversion of
the large (of the order of the number of pixels in the map) matrix
$P^{T}\textbf{N}^{-1}P$, which is the covariance matrix of the GLS
estimator \citep{Lupton93}. One key simplifying assumption is to
invoke that the noise is stationary. In this case, the $\textbf{N}$ matrix has
a Toeplitz form which can be approximately treated as circulant,
ignoring boundary effects \citep{Natoli01}. A circulant matrix is
diagonal in Fourier space and its inverse is also circulant, so the
product between $\textbf{N}^{-1}$ and a vector is a convolution
between the same vector and a filter provided by any of the rows of
the matrix. In the following we will refer to any of these rows as a
noise filter. Its Fourier transform is the inverse of the noise frequency power
spectrum.  

Considering the conditions listed above, the GLS map making algorithm
performs the following operations, starting with rewriting the Equation \ref{eq:GLS} in the form 

\begin{equation}\label{eq:gls_rewrite}
(P^{T}\textbf{N}^{-1}P)\textbf{m}_{0}-P^{T}\textbf{N}^{-1}\textbf{d}=\textbf{r}
\end{equation}

where $\textbf{m}_{0}$ is the starting map used at the first iteration, generally the naive map. 

The $P \textbf{m}_{0}$ product projects the map onto a timeline.  Application of $\textbf{N}^{-1}$: this is a convolution which can be performed in Fourier space. Application of $P^T$: this step projects the convolved timeline back into a map. 

The second term performs the convolution with the filter (applying  $\textbf{N}^{-1}$ to the data vector $\textbf{d}$ in Fourier space) and then the projection of the convolved timeline into a map (applying $P^T$ to the product $\textbf{N}^{-1}\textbf{d}$). 

Then, we need to evaluate the residual \textbf{r}. If the residual is higher than a fixed threshold, it is used to produce a new map, $\textbf{m}_{1}$, as described in \citet{CG}. This map will be considered instead of $\textbf{m}_{0}$ for evaluating again the Equation \ref{eq:gls_rewrite}, until convergence. This is achieved by running a Conjugate Gradient algorithm, an iterative
  method useful to obtain a numerical solution of a system \citep{CG},
  until convergence is reached with the residual lower then the threshold.

The algorithm outlined is the same described in \citep[``unroll, convolve and bin":][]{Natoli01} and is implemented in the ROMAGAL
code. The next section explains the strategy employed to estimate the
noise filters used by ROMAGAL directly from in-flight data. 

%========================================================
\subsection{Noise estimation}\label{sec:noise_filters}

In order to estimate the noise filters for ROMAGAL we need to
investigate the noise statistical properties in the timelines. Data
are mostly affected by two kind of statistical noise: $1/f$ noise due
both to the electronics and thermal background radiation from the telescope
or the instruments, and photon noise (see \citealt{Poglitsch08,
  Schulz08}).  

The detector $1/f$ noise arises in the electronic chain and it impacts
particularly regions with low signal-to-noise ratio (SNR), where only
diffuse emission is present. In those regions it can be of the same
order of magnitude of the signal or even higher. In these cases GLS
treatment is particularly effective.  

Photon noise is due to statistical fluctuation in the photon
counts. This process follows poissonian statistic, so the SNR is
proportional to the square root of counts. Since poisson distribution
tends to be Gaussian for large numbers, we can approximate photon
noise as Gaussian on the map if the number of counts is large
enough. %This imposes a constraint on the final map pixel size, as
%discussed in the next section. 
 
Since bolometers are organized in matrix and sub-matrix, the signal of a bolometer can be correlated with the signal of another, generally adjacent, bolometer. These effects could be both
statistical and deterministic. We already described how to remove the deterministic common mode from
TOD (like the thermal bath variations, see Section \ref{sec:preprocessing}).

One possible source of statistical cross-correlated noise is the crosstalk between bolometers: the signal of
one pixel may contaminate the signal of its neighbors through the
capacitive inductive coupling, generating a common mode called
``electrical crosstalk''. On the contrary, ``optical crosstalk'' is
due to diffraction or aberrations in the optical system which could
drive an astronomical source to fall on inappropriate detectors
\citep{Griffin09_pipeline}. 

We then analyze the residual contribution of the statistical component
of the correlated noise. We found that the residual correlated noise
level in each pixel is negligible with respect to the intrinsic
detector noise level for both PACS and SPIRE instruments, as described
in the following. 

In principle, the noise
properties vary significantly across the array and we had to estimate
the noise power spectrum for each bolometer. 
To do that we have processed ``blank" sky mode (i.e. filled with
negligible contribution from sky signal) data acquired during the
PV Phase. 
 
\begin{figure}
\centering
\includegraphics[width=8cm] {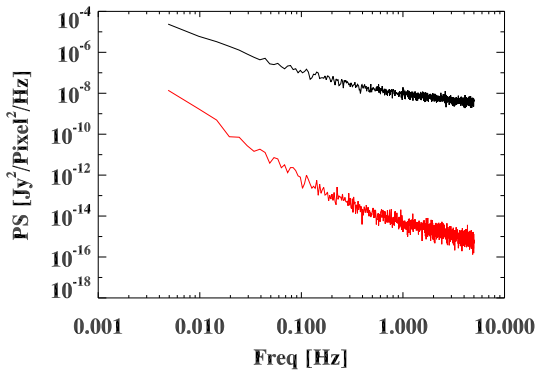}%{./Img/blue_noise_spectrum.eps}
\caption{Black line: typical noise spectrum of a PACS $160$\mm\ detector,
  estimated on blank sky data. Red line: cross spectrum between two
  detectors of the same subarray. The level of the cross-correlated noise is
  significantly under the noise level of each single bolometer, so we
  can reasonably neglect it.} 
\label{fig:PACS_blue_noise_spectrum}
\end{figure}

In Figure \ref{fig:PACS_blue_noise_spectrum} we show a typical noise spectrum
estimated for a pixel of the $160\mu$m  PACS band (black) and the
cross-spectrum between two adjacent bolometers (red). The
cross-spectrum evaluates the impact of the cross-correlated noise in
the frequency domain between two different bolometers. The level of
the cross-correlated noise is at least 4 order of magnitude below the
auto-correlated noise power spectrum of each pixel. Note that this means we do not see any relevant cross-correlated noise, despite the fact that crosstalk can be present into the timeline. 

\begin{figure}
\centering
\includegraphics[width=8cm] {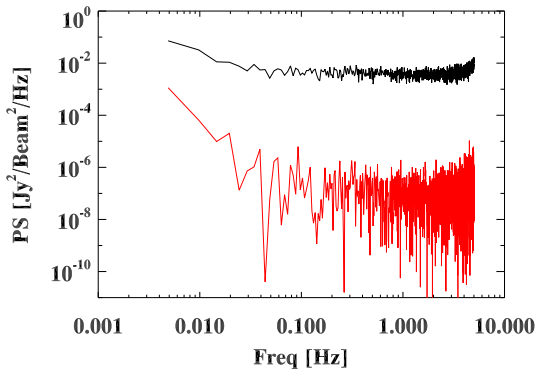}%{./Img/PSW_noise_spectrum.eps}
\caption{Same as Figure \ref{fig:PACS_blue_noise_spectrum} for a SPIRE
  $250$\mm\ bolometer. For SPIRE also the noise level of cross-spectrum is
  reasonably negligible with respect to the auto spectrum level.} 
\label{fig:SPIRE_psw_noise_spectrum}
\end{figure}

In Figure \ref{fig:SPIRE_psw_noise_spectrum} we
show noise power spectra of the 250$\mu$m SPIRE band bolometers. 
Also in this case the cross-spectrum is negligible. 
 
Noise spectra of both PACS and SPIRE display low-frequency noise
excess ($1/f$). In case of SPIRE spectra (Figure
\ref{fig:SPIRE_psw_noise_spectrum}) a high
frequency rise is also evident, which is due to the deconvolution of the bolometric
response function. PACS spectra do not show this behavior because the
bolometer transfer function is not deconvoluted by the standard
pipeline. 


%========================================================
\subsection{From ROMA to ROMAGAL}\label{sec:roma_romagal}
 
The ROMAGAL GLS code has been optimized to recover the features in the
\HG\ fields with high accuracy. 
 
Hi-GAL observes the Galactic plane where the dynamic range of sources
spans over several orders of magnitudes. This poses strong constraints
on the map making algorithm: both the weak diffuse emission and the
bright compact sources in e.g., star forming regions have to be
recovered with high accuracy. 
The signal often exhibits steep gradients that are hard to follow for
the GLS solver, which relies on the assumption that the sky signal
does not vary significantly within a resolution element (see below
\ref{sec:signal_striping}). At the same time, several systematics
affect the dataset. As explained above, many of them are cured at the
preprocessing level. However, their removal generates a conspicuous
amount of transient flagging, that must be correctly handled by the
GLS code. 
 
The core of ROMAGAL is the same of the ROMA code \citep{deGasperis05}
where the input-output routines have been deeply modified to adapt to
the HIPE generated dataset. ROMAGAL inputs are the TOD generated by
HIPE, pointing and transient flagging. These have the same format for
both PACS and SPIRE. ROMAGAL outputs are fits file containing the
optimal map, arranged in standard gnomonic projection routines. 
The code is written in FORTRAN 95 and relies on the MPI library for
parallel calls. It runs on the \HG\ dedicated machine, called BLADE,
a cluster of 104 processors at 2.5 GHz each and 208 Gb RAM total. Its
nodes are interconnected with MPI-infiniBAND. The machine is located
at IFSI-Rome. 

As explained in the previous Section, the computation of Equation~\ref{eq:GLS} is unfeasible due to the size of the map's covariance
matrix. However, we assume the noise of each \HG\ field to be
stationary to set up an FFT based solver built upon a conjugate
gradient iterative algorithm (see Section \ref{sec:mapmaking}). Such
a scheme can estimate the final maps with a precision of order of
$\epsilon=10^{-8}$ in $\sim$ 150 iterations for \HG\ . ROMAGAL
computational time scales linearly with the size of the dataset and
only weakly with the number of pixels in the output maps. The scaling
with the number of processors is highly optimal in the range of cores
required for the \HG\ analysis ($< 50$). For the largest channels
(PACS blue band), a final GLS map of a $2^\circ \times 2^\circ$ field requires about 16 Gbytes of RAM and 1400 sec on 8
cores.  Due to the high number of array pixels (2048), this channel is
the largest dataset to analyze as well as the most demanding in terms
of computational resources. Further information on resource
consumptions can be found in Table \ref{tab:computing_resources}. 

\begin{table}
\begin{center}
\begin{tabular}{|c|c|c|}
\hline  \textbf{Band} & \textbf{Total Time (sec)}  & \textbf{RAM (Gb)}\\ 
\hline  $70\mu$m &  $\sim$ 1400 & 16\\ 
\hline  $160\mu$m &  $\sim$ 1000 & 8\\ 
\hline  $250\mu$m & $\sim$  180 & 4\\ 
\hline  $350\mu$m &  $\sim$ 130& 1\\ 
\hline  $500\mu$m &  $\sim$ 100& 1\\ 
\hline 
\end{tabular} 
\end{center}
\caption{Time and minimum RAM amount required from ROMAGAL for each
  PACS and SPIRE band using 8 BLADE processors.} 
\label{tab:computing_resources}
\end{table}

%========================================================
\subsection{Optimal treatment of transient flagging}

As mentioned above, the timelines are inspected for bad data samples
that must be excluded from map making as part of the preprocessing
pipeline. Bad data can arise due to a variety of reasons. They are
generally caused by transient events, either unexpected (e.g.,
glitches, anomalous hardware performance) or expected (e.g.,
detectors saturating because of a bright source, observation of a
calibrator).    
Once identified, a flag record is generated and stored for these
anomalous events, so that their contribution can be safely excluded
from map making. Flags in the TOD pose a potential problem for ROMAGAL
because its solver is based on the FFT as discussed in the previous
section. The FFT requires the timeline to be continuous and uniformly
sampled. Since noise in the PACS and SPIRE data is correlated, just
excising the flagged sampled to fictitiously create a continuous
timeline would interfere  with noise deconvolution, and is thus not a
safe option. Instead, we advocate using a suitable gap-filling
technique. The rest of this section is mostly devoted to defining
which, among the various options for gap filling, is best suited for the
\HG\ maps.  

It is important to realize that the output map will depend on the
content of the flagged sections of the TOD even if these values are
not associated to any (real) map pixel. This is due to the convolution
performed within the solver that ``drags'' out data from the flagged
section of the timeline, even if, when the data are summed back into a
map, the $P$ operator is applied only to the unflagged samples. Since
one is not in control of the content of the flagged section of the
timeline, a kind of gap-filling must be performed in any case. We have
tested different recipes for gap-filling making extensive use of
simulations. We have treated separately the signal and noise
components of the timelines, running separately noise dominated and
signal dominated cases, because the behavior towards flags of the two
components is different, as it will be shown below. 

\begin{figure*}
\centering
\includegraphics[width=10cm] {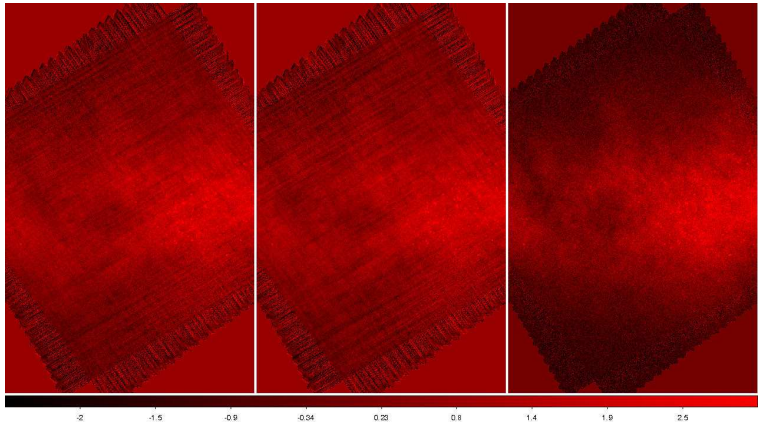}
\caption{Shown are the results obtained in a noise
dominated regime (normal \HG\ noise is amplified by a factor
100). Left panel is the map obtained replacing flagged data samples
with null values (clearly it does not work), middle panel is the map
obtained replacing data samples with unconstrained noise realization
(does not work either), right panel shows the map obtained using our
NCR code (does work). 
\label{fig:zeroflags_noiseonly_noCNR}}
\end{figure*}

The simplest form of gap filling is to remove the content of the flags
altogether, replacing the flagged sections with a nominal constant
(usually null) value. This works very well on a signal-only simulation
of the Hi-GAL field. However, it fails dramatically when noise is
present, as evident from Figure  
\ref{fig:zeroflags_noiseonly_noCNR} (left panel), where a markedly
striped pattern in the reconstructed map is seen (in this simulation,
the \HG\ noise has been amplified by a factor 100 to make its
pattern more evident). The reason for this failure is readily
understood: The GLS map making employed requires noise stationarity
(see Section  \ref{sec:roma_romagal} above), which is obviously not
preserved by zeroing the gaps. A less obvious effect is that even if
gaps are filled with a noise realization with the correct statistical
properties, but unconstrained, the GLS map making is bound to fail as
well, as shown in the middle panel of Figure
\ref{fig:zeroflags_noiseonly_noCNR}. A noise realization is said to be
constrained when it respects certain boundary conditions
\citep{1991ApJ...380L...5H}, which in our case are represented by the
noise behavior outside the gap. Unconstrained noise inside the gap,
despite having the correct statistical properties, creates a border
discontinuity that causes the GLS map maker to behave sub-optimally
\citep{2002PhRvD..65b2003S}. We have employed a code to create
Gaussian noise constrained (NCR) realizations, based on the algorithm
set forth by \citet{1991ApJ...380L...5H}. The code uses in input the
noise pattern and statistical properties, as measured from the
timelines. The results on noisy simulations are excellent, as set forth
by the third (rightmost) panel in Figure
\ref{fig:zeroflags_noiseonly_noCNR}. Note, however, that Figure
\ref{fig:zeroflags_noiseonly_noCNR} refers to a noise dominated
simulation. We now turn to discuss the effect of a non negligible
signal contribution (a far more realistic case for \HG\ ). 

We have verified that the presence of non negligible signal in the
timelines does not affect the results provided that the NCR is
performed using the underlying noise field as a baseline. Measuring
the latter in presence of signal is however impractical. It would be
significantly simpler if the NCR could be run directly on the
timelines themselves, constraining thus the (fake) noise realization
within the gap to the outside noise plus signal (true)
values. Unfortunately, this poses a problem for \HG\ data: large
signal jumps are present in the field and the resulting gap filling
realizations are affected in a non negligible manner by the boundary
conditions, at least with the present version of ROMAGAL. This
behavior is different from what happens for experiments aimed at the
Cosmic Microwave Background (see e.g. \citet{Masi06})
where NCR codes are routinely run on the timelines as they are. In
order to find a workaround that would spare us the inconvenience of
estimating the underlying noise field to serve as a NCR input, we have
modified the flag treatment of the ROMAGAL code as explained in the
following. 

\begin{figure}
\centering
\includegraphics[width=8cm] {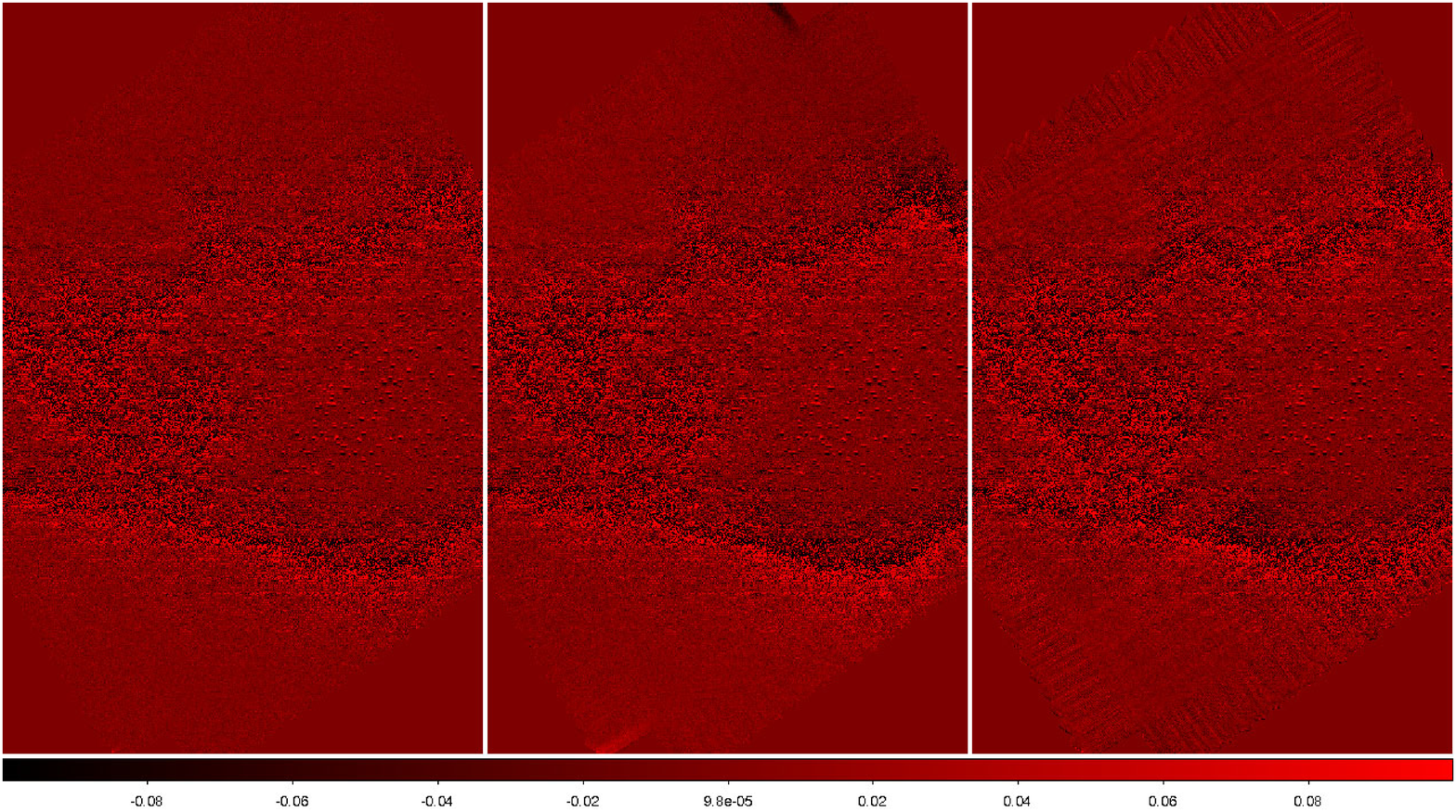}
\includegraphics[width=8cm] {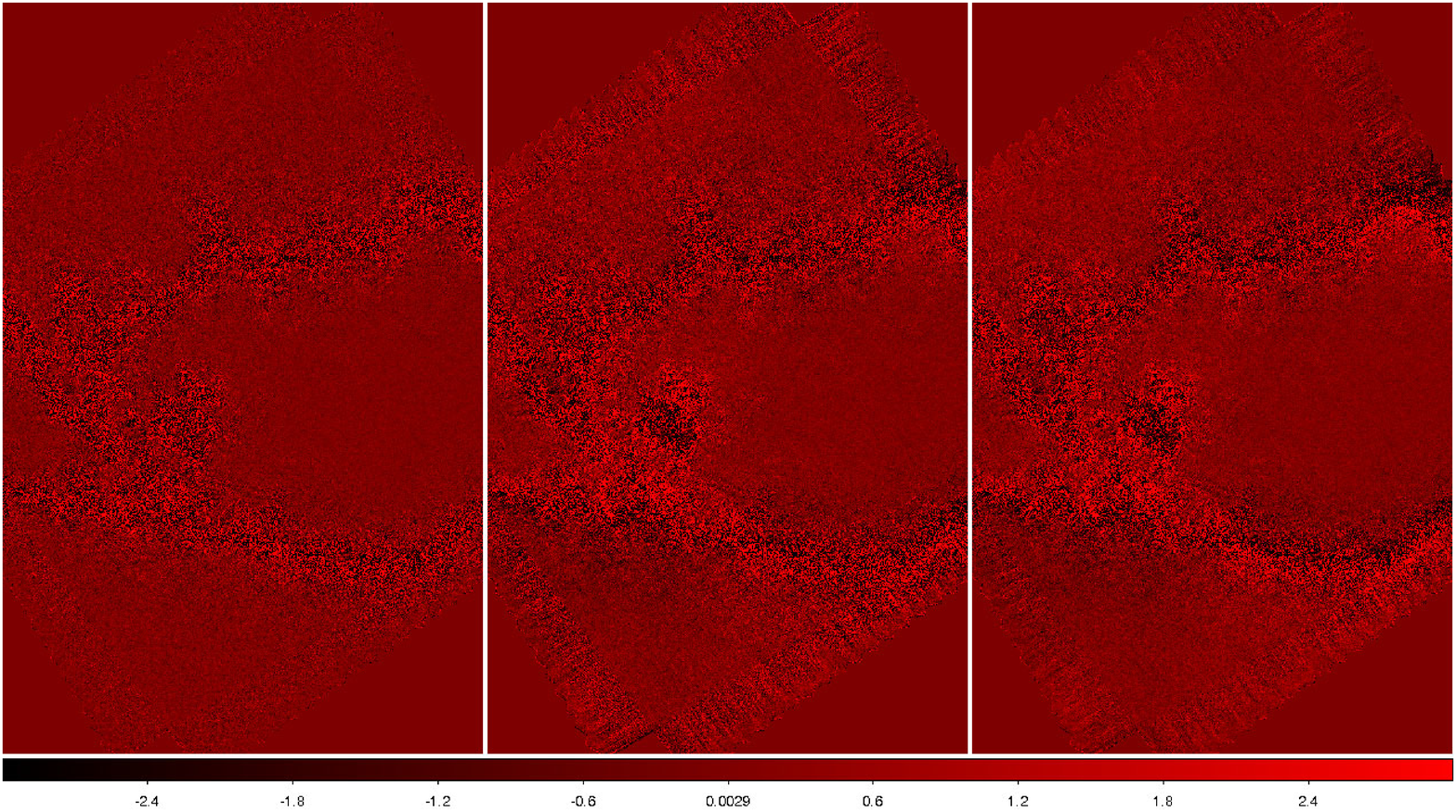}
\caption{Top row: For a signal dominated case, shown are the relative
  differences between the input simulated map and the output obtained
  with ROMAGAL (in the ``virtual map'' mode) with NCR performed on the
  underlying noise (left), on the original signal plus noise TOD
  (middle), and without NCR after replacing the gaps with null
  values. The latter case is clearly striped, but no signal related
  artifacts are present. Bottom row: For a noise dominated case, shown
  again are the relative difference versus the input map obtained by
  ROMAGAL with ``virtual map'',  assuming NCR on underlying noise only
  (left), without NCR (middle) and with NCR on the signal plus noise
  TOD (right). As expected, the left panel shows the best residuals,
  but the right one appears as a good compromise (see also text). 
\label{fig:reldiff_realcnr_nvirtcnr_zeroflags_vs_input}}
\end{figure}

The original version of ROMAGAL makes use of a single extra pixel
(dubbed ``virtual pixel'') to serve as junk bin where the contents of
the gaps are sent when applying the $P^T$ operator within the ROMAGAL
solver. This approach, as stated above, works excellently in presence
of both signal and noise, irrespective of their relative amplitude,
provided the NCR code assumes the underlying noise field as a baseline
to perform the realization. In order to relax this assumption, we have
modified ROMAGAL to take into account not a single virtual pixel but
an entire virtual map. In other words, we introduce a virtual
companion for each pixel of the map, and use it as a junk bin to
collect the output from the gaps they correspond to. The hope is to
redistribute the content of the flagged sections more evenly,
preventing artifacts. This approach obtains satisfactory results when
the NCR code is run on the original (signal plus noise) timelines, as
shown in Figure \ref{fig:reldiff_realcnr_nvirtcnr_zeroflags_vs_input}.

To summarize our findings:
\begin{itemize}

\item Using a NCR realization code is {\it always} necessary in order to avoid artifact striping effect into the map

\item If the NCR code is run on the underlying noise-only timelines
  (which are cumbersome to estimate) we obtain the best quality output
  map, with no signal related artifacts and a noise standard deviation
  which is lower (by a factor $\sim 2$) with respect to the case in
  which no NCR is performed. 
\item Running the NCR on the original timelines is possible with the
  ``virtual map'' approach. No signal artifacts are detected in the
  difference maps and the advantage in terms of noise standard
  deviation with respect to the no NCR case is still present, and of
  order 10\% to 20\% on average. 
\end{itemize}
We have therefore chosen this  latter approach as the baseline for the \HG\ pipeline. 

%========================================================
\section{ROMAGAL maps}\label{sec:SDP_results} 

In this section we analyze the final map obtained running our dedicated pipeline. We analyze the Point Spread Function (PSF) for the five bands in order to fix the resolution of the ROMAGAL maps. We compare the final GLS map with the naive map and we
point out the differences and the capability to recover the diffuse
emission from the GLS map. Finally we discuss about the noise
residuals on the maps. 

%========================================================
\subsection{Point Spread Function and pixel size}\label{sec:PSF}

The angular resolution (pixel size) of the final map is a free
parameter. Its choice is a compromise between competing
requirements. A small pixel size assures a correct sampling of the PSF;

indeed, assuming a Gaussian profile for the PSF (which is reasonable as discussed in the following), the Nyquist theorem imposes that to better sample a 2-d image we need
to set a pixel size which is at most one third of its FWHM value.  

On the other hand, a too small pixel size can cause the loss of
redundancy, useful to reduce the white noise level, and (even) some
non-observed pixels in the final map. 

The diffraction limited beam of PACS at $70\mu$m is $5.2\arcsec$. Thus,
we should build the map with a pixel size of at least $1.8\arcsec$. However, due to the limited bandwidth available for
transmission, especially in PACS/SPIRE Parallel mode, PACS frames are coadded on-board the satellite before broadcasting. For
the $70$\mm\ channel, a coaddition of 8 consecutive frames is applied by
on-board software.  Since the acquisition rate is 40Hz and the scanning
speed for \HG\ is set to $60\arcsec /s$, two close frames are
snapshots of the sky acquired $1.5\arcsec$ apart. Due to coaddition,
the satellite provides one averaged frame every $12\arcsec$; in
spatial coordinates, this is twice the beam width of the PACS blue channel. 
The measured PSF then is not the diffraction limited beam but it
results in an elongation along the scan direction due to the
convolution of the coaddition with the beam. As shown in \citet{Lutz09},
the observations of Vesta and $\alpha$Tau with the blue channel
evidenced a FWHM equal to $5.86\arcsec \times12.16\arcsec$ as a result
of a 2-d Gaussian fitting, elongated in the scan direction.  

PACS $160\mu$m is also affected by the averaging on-board, but only 4
 frames are coadded together. The nominal instrumental beam is
 $12.0\arcsec$, while the measured is $11.64\arcsec\times15.65\arcsec$
 \citep{Lutz09}, elongated along the scan direction. However, in this
 case we can sample the beam without issues, and the effect of coaddition
 on the final map is negligible.
 
For the \HG\ fields the scanning strategy consists of two orthogonal AORs, therefore the redundancy regularizes the
PSF, resulting in approximately 2-d symmetric
Gaussian profile, as shown in Table  \ref{tab:time_beam_pixel}.  

According to the values reported in  Table  \ref{tab:time_beam_pixel}, we observe
 quasi-symmetric beams with an averaged
ellipticity of less than $15\%$ for both blue and red channel and the
axis oriented randomly with respect to the scan direction.

We choose a pixel size of $3.2\arcsec$ for the PACS $70\mu$m
band which samples the observed beam at Nyquist frequency. Below this threshold, the diffuse emission
areas become too noisy due to the low SNR.
Similarly, we can choose a pixel size of
 $4.5\arcsec$ for red band without loosing redundancy.  

SPIRE does not suffer from on-board coadding and the detectors were
built to reach the band diffraction limit. In-flight data show
that the SPIRE beam is well approximated by a 2-d asymmetric Gaussian
curve with the major axis orientation independent of the
scan direction, with an ellipcticity not bigger then $10\%$ (see \citet{Sibthorpe10}). We set the pixel size for
each SPIRE band equal to one third of the nominal beam. 
In Table \ref{tab:time_beam_pixel} we also report the beam measured in
the SPIRE maps. 

The average ellipticity we observe agrees with found by \citet{Sibthorpe10}. On the contrary, while the FWHM for the two axis found by \citet{Sibthorpe10} are in agreement with the nominal (within the error), our measured beam results in a FWHM larger then the nominal of $\sim25\%$. 

\begin{table*}
\begin{center}
\begin{tabular}{|c|c|c|c|c|}
\hline  \textbf{Band} & \textbf{Nominal Beam (arcsec)}  & \textbf{Measured Beam (arcsec)} & \textbf{Ellipticity} & \textbf{Pixel Size (arcsec)}\\ 
\hline  $70\mu$m & $5.2\times 5.2$ & $\sim 9.7\times \sim 10.7$ & 14.6\% &  3.2 \\ 
\hline  $160\mu$m &  $12.0\times 12.0$ & $\sim 13.2\times \sim 13.9$ & 14.7\%& 4.5\\ 
\hline  $250\mu$m &  $18\times 18$ & $\sim 22.8\times \sim 23.9$ & 8.3\%& 6.0\\ 
\hline  $350\mu$m &  $24\times 24$ & $\sim 29.3\times 31.3$ & 8.8\%& 8.0\\ 
\hline  $500\mu$m &  $34.5\times 34.5$ & $\sim 41.1\times \sim 43.8$ & 9.7\%& 11.5\\ 
\hline 
\end{tabular} 
\end{center}
\caption{Nominal (2nd column), map-measured beam (two AOR, 3rd column) and ellipticity (4th column) of each band.} 
\label{tab:time_beam_pixel}
\end{table*}

%========================================================
\subsection{Hi-GAL SDP results}

The quality of the outcome can be qualified by the comparison between the igls maps with the naive maps. In fact, the naive map is the simple averaging of signal recorded by every spatial pixel and it represents ``the least compromised view of the sky". 

Since the TOD are created at the end of the preprocessing steps, when the data are a combination of only signal and $1/f$ noise, we expect the $1/f$ residuals in the naive map as well as a ``pure" sky map produced by ROMAGAL. 

\begin{figure}
\centering
\includegraphics[width=8cm]{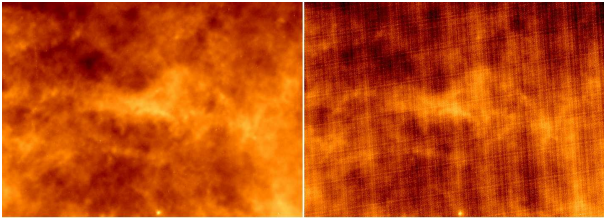}
\caption{Left: particular of the ROMAGAL map of the red PACS array,  \30 field. Right: particular of the naive map, same
  band and field. The $1/f$ noise is evident on the naive map, as well
  as its minimization is evident on the GLS map without loosing the
  signal of the diffuse component.} 
\label{fig:PACS_l30_igls_vs_bin}
\end{figure}

In Figure \ref{fig:PACS_l30_igls_vs_bin} a comparison between the
naive map and the ROMAGAL map of \30 field at $70\mu$m is shown. The
GLS code is capable to remove the $1/f$ residuals without loosing any
signal both on bright sources and on diffuse emission. 

 In particular, we choose three main proxies:

\begin{itemize}
\item   the difference between naive and igls should show only a pattern due to the $1/f$ noise residuals in the binned map. The pattern of this low-frequency noise is recognizable as stripes superimposed on the sky signal in the naive map. The stripes are the consequence of the $1/f$ noise due to the adopted scanning strategy.\\
In Figure \ref{fig:diff_PACS_l30_red_igls_bin} we show the difference
between igls map and naive map. The $1/f$ noise is removed in the igls
map but not in the naive. The residual stripes due to the
low-frequency noise are clearly visible. 

\begin{figure}
\centering
\includegraphics[width=5.25cm]{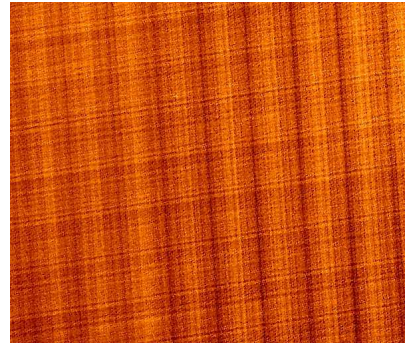}
\caption{Particular of map difference between PACS $160\mu$m igls and naive, same region of the previous Figure. The
  stripes due to $1/f$ removal made in the igls are evident.} 
\label{fig:diff_PACS_l30_red_igls_bin}
\end{figure}

\item the source fluxes should be the same between the igls and naive maps.\\
 This item is quantified by the map difference, where the pattern is due only to the noise without any residual signal, except across the very brilliant source. This last effect is discussed on the next Section. %..................................

\item a statistical analysis of the background noise level should show a decrease of the rms value in the igls with respect to the naive map.\\
In Tables
\ref{tab:rms_PACS} and \ref{tab:rms_SPIRE} we report the rms residuals of the PACS and SPIRE maps respectively, calculated
in a diffuse emission area of each map. Since the flux between the maps is conserved, a decreasing of the rms noise level assures an increasing of S/N ratio in the ROMAGAL maps.

\begin{table}
\begin{center}
\begin{tabular}{|c|c|c|c|}

\hline
\multicolumn{4}{|c|}{\textbf{PACS \30 field}}\\
\hline \textbf{Band}&  \textbf{rms igls (MJy/pixel)}& \textbf{rms naive(MJy/pixel)}& \textbf{ratio}\\ 
\hline  $70\mu$m &  0.0085&  0.026& $\sim$ 3.1 \\ 
\hline  $160\mu$m &  0.047&  0.102&  $\sim$ 2.2\\ 
\hline 
\hline
\multicolumn{4}{|c|}{\textbf{PACS \59 field}}\\
\hline \textbf{Band}&  \textbf{rms igls (MJy/pixel)}& \textbf{rms naive(MJy/pixel)}& \textbf{ratio}\\ 
\hline  $70\mu$m &  0.004545&  0.02208& $\sim$ 4.9 \\ 
\hline  $160\mu$m &  0.01899&  0.03586&  $\sim$ 1.9\\ 
\hline 
\end{tabular} 
\end{center}
\caption{rms compared from GLS and naive map on both the SDP observations,
  for PACS bands, measured on a background region of $50\times$50
  pixels. In the last column is reported the ratio between the naive
  and GLS rms.} 
\label{tab:rms_PACS}
\end{table}

The ratio between the naive and igls rms shows an improvement of a factor $\sim 2-5$ in the PACS ROMAGAL maps, and a factor $\sim 1-2$ in the SPIRE case. The difference is mostly due to an intrinsically different $1/f$ noise level.

\begin{table}
\begin{center}
\begin{tabular}{|c|c|c|c|}
\hline
\multicolumn{4}{|c|}{\textbf{SPIRE \30 field}}\\
\hline \textbf{Band}&  \textbf{rms igls (MJy/beam)}& \textbf{rms naive(MJy/beam)}& \textbf{ratio}\\ 
\hline  $250\mu$m &  0.1749&  0.2868& $\sim$ 1.6 \\ 
\hline  $350\mu$m &  0.1569&  0.2302&  $\sim$ 1.5\\ 
\hline  $500\mu$m &  0.2659&  0.4065&  $\sim$ 1.5\\ 
\hline 
\hline
\multicolumn{4}{|c|}{\textbf{SPIRE \59 field}}\\ 
\hline \textbf{Band}&  \textbf{rms igls (MJy/beam)}& \textbf{rms naive(MJy/beam)}& \textbf{ratio}\\ 
\hline  $250\mu$m &  0.09857&  0.1123& $\sim$ 1.1 \\ 
\hline  $350\mu$m &  0.0734&  0.08164&  $\sim$ 1.1\\ 
\hline  $500\mu$m &  0.1073&  0.2101&  $\sim$ 1.9\\ 
\hline 
\end{tabular} 
\end{center}
\caption{rms compared from GLS and naive map on both the SDP observations,
  for SPIRE bands, measured on a background region of $50\times$50
  pixels. In the last column is reported the ratio between the naive
  and GLS rms. The $1/f$ noise is less evident in the SPIRE bolometers
  with respect to the PACS one, but its effect is still remarkable.}
\label{tab:rms_SPIRE}
\end{table}

\end{itemize}

%========================================================
\subsection{Consistency of data analysis assumptions}\label{sec:signal_striping}

One of the assumptions of ROMAGAL, as well as of all Fourier based GLS map making codes, is that the underlying sky signal is constant within a chosen resolution element. If this is not the case, artifacts (stripes) will be generated in the final map, contributing to the so called pixelization noise \citep{Poutanen06}. In the case of Hi-GAL the situation is complicated by several effects: 

\begin{itemize}

\item On-board coaddition of samples: each PACS 70$\mu$m (160  $\mu$m) frame is the result of an on-board average of eight (four)
consecutive frames  reducing the effective sampling frequency of the instrument (see Section \ref{sec:PSF}). Thus, sky signal is low-pass filtered by only partially effective data-windowing, rather then telescope response, leaving room for signal aliasing.

%at high frequency is not properly smeared by the telescope's response at the native sampling rate, but rather by less effective data windowing, leaving room for signal aliasing. 

The map making code is quite sensitive to aliasing since it works in Fourier space. The situation is worsened by the large dynamic range of the Hi-GAL fields, especially when scanning across bright sources.

\item Time bolometer response induces signal distortions along the scan. While within HIPE the SPIRE detector response is deconvoluted from the data \citep{Griffin09_pipeline}, the same is not true for PACS. Redundancy in each pixel is obtained by scans coming from different directions, thus the effect contributes further signal mismatch at the pixel level.

\item Pointing error: as analyzed in detail in \citet{Herschel} the pointing performance of Herschel, which mean the capability of assign coordinates to each sample in a given reference frame, can be affected by several pointing error effects; the main contributor is due to the angular separation between the desired direction and the actual instantaneous direction, based on the position-dependent bias within the star-trackers. 

The Herschel AOCS goal is that the mismatch between real coordinates and assigned
coordinates along the scan-leg is smaller than $1.2\arcsec$ at 1 sigma \citep{Herschel}. So that a 2$\sigma$ event becomes significant compared to the PSF of the PACS blue band.

\end{itemize} 

\begin{figure}
\centering
\includegraphics[width=6cm]{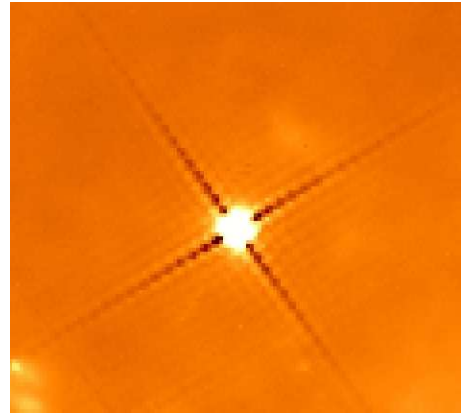}
\caption{Zoom around a compact sources for the PACS blue optimal
  map. The striping dragged along the scan directions are
  remarkable.}  
\label{fig:crosses_blue}
\end{figure}

All of the above effects challenge the basic assumptions (no signal aliasing and no sub-pixel effect) under which ROMAGAL works.  Our simulations suggest that signal aliasing contribute significantly more than the other two. The net result on maps is the striping of bright sources in the GLS maps. An example is shown in Figure \ref{fig:crosses_blue} for the PACS blue band.

It is important to notice that the stripes are present only around the point-like sources (where - of course - the signal aliasing is more evident), regardless of their magnitude. However, magnitude influences the amplitude of the stripes. For a source peak which is only 10 times higher that the rms background, the intensity of the stripes are within $1\sigma$ from the background dispersion. For the more intense sources, the stripes magnitude in the pixels surrounding the PSF shape can be $100\sigma$ times away from the background value.

Since these stripes are produced within the GLS solver, which performs repetitive convolutions along the scan directions, but do not affect the naive map the obvious workaround is to implement dedicated map making which considers a different analysis around the bright sources.
 
 However, the detailed accounting of the above effects and the enhanced map making strategies to address them  will be the subject of a forthcoming paper \cite{Piazzo11}.

%========================================================
\section{Summary and conclusions} \label{sec:Conclusions}

This paper describes in detail all steps of the processing of Herschel data, from originally downloaded frames to high-quality maps, used in the framework of Hi-GAL survey. Hi-GAL data are taken in fast scan mode ($60\arcsec$/sec ) and simultaneously by PACS and SPIRE (Parallel mode). We test our pipeline reducing data from the Science Demonstration Phase and present results, taking as proxy for the quality of the final images their comparison with naive maps. 

We divided data processing into two distinct phases: preprocessing and mapmaking. Pre-processing aims to accurately remove systematics and random effects in the data, in order to prepare them for the ROMAGAL map making algorithm, which implements the minimum variance GLS approach in order to minimize the noise into the data. It turns out that NCR is a fundamental step in the pre-processing because ROMAGAL, as an FFT mapmaking code needs continuous and regular data time series as input.

Noise residuals in the diffuse emission of the two test fields (SDP Hi-GAL data, two 2$^{\circ}\times2^{\circ}$ tiles centered on the Galactic plane at l = 30$^{\circ}$ and l = 59$^{\circ}$) show that we obtain optimal maps, getting rid of glitches and systematic drifts, as well as minimizing the 1/ f and white noise component. The remaining effects, which do not affect the overall quality of the maps except across the bright source on the PACS 70$\mu$m maps, are under investigation that will appear in a dedicated publication to be available shortly. 

%========================================================
\section*{Acknowledgements}
Special thanks to G\"oran Pilbratt for allowing the using of PV Phase
blank data. 

\bsp

\bibliographystyle{mn2e}
\bibliography{/Users/alessio/Work/Papers/MyPap/bibliography.bib}

\label{lastpage}

\end{document}